\documentstyle[12pt]{article}
\begin{document}
\textwidth=18cm
\textheight=22cm
\begin{center}{\Huge {\bf Nucleation in Ising ferromagnet by a 
field spatially spreading in time}}\end{center}

\vskip 0.5cm
\begin{center}{\it Muktish Acharyya}\\
{\it Department of Physics, Presidency University}\\
{\it 86/1 College Street, Calcutta-700073, India}\\
{\it E-mail:muktish.physics@presiuniv.ac.in}\end{center}

\vskip 1cm

\noindent {\bf Abstract:} The nucleation in Ising ferromagnet has been studied
by Monte Carlo simulation. {\it Here, unlike the earlier studies, 
the magnetic field is spreading over the
space in time.} The nucleation time is observed to increase as compared to that
in the case of static field. The clusters of negative spins is observed to grow
from the center. The growth of effective magnetisation is 
studied with temperature
and the strength of spreading magnetic field. The ratio of nucleation time and
effective time is also studied with strength of spreading magnetic field. 
{\it The effective time would introduce itself as a new scale of time
in the case of nucleation by spatially spreading magnetic field.}

\vskip 2cm



\newpage

\noindent {\large {\bf I. Introduction:}} 

The dynamical aspect of Ising ferromagnet, 
particularly concerned with the nucleation and growth of clusters, 
has become an active area of theoretical research
over a very long period of time\cite{rik0,rik1}. 
The lifetime of metastable states and its dependences on the temperature
and applied magnetic field are main focus of attention. Extensive Monte Carlo studies are performed
in last few decades\cite{mads,rik2,nowak}. The large scale simulational results\cite{mads} 
show strong agreement with classical nucleation theory.
Some of the important studies may be mentioned briefly as follows: classical nucleation theory
was reexamined by extensive Monte Carlo simulations and connection with the hysteresis was found
\cite{mads}. In this case, the applied magnetic field was taken uniform over the space and constant
in time. Later, the nucleation in Ising ferromagnet was studied in presence of a time dependent
magnetic field to analyse the hysteresis phenomenon. However, the field was uniformly distributed
over the space\cite{rik1}. 
The magnetisation reversal was studied in Ising ferromagnet by a pulsed field\cite{arko}.
The magnetisation reversal was also studied in Ising ferromagnet by a periodic
impulsive field \cite{magrev}. Magnetisation switching behaviour was also
studied \cite{nowak} in vector spin model with long range 
dipolar interaction by Monte Carlo simulation.
The magnetisation reversal and switching were studied experimentally\cite{expt1,expt2}. 
It may be noted here all theoretical and experimental 
investigations are made in the presence of either
a static magnetic field or by a field varying in time\cite{rik2}. In this context, it
should be mentioned that, the recent Monte Carlo study\cite{rik3}
 yields a heat assisted magnetisation
reversal by a decaying and spatially spreading temperature pulse. This study is technologically
important for high density ultrafast magnetic type recording.
{\it But in the earlier studies on the magnetisation reversal by time dependent
magnetic field, the field was distributed uniformly over the space}. 
No such study was found where the
spatio-temporal variations of field was considered, in the case of nucleation in Ising
ferromagnet. It is interesting to know how the magnetsation reversal
happens when the field is spreading spatially in time.

In the present paper, the nucleation in the Ising ferromagnet was studied where the field is spreading
spatially in time. The Monte Carlo simulation with Metropolis single spin flip dynamics
\cite{binder} is used here.
In section-II the model and the simulation scheme are described. The simulational results are 
reported in section-III. The paper ends with summary in section-IV.

\vskip 1cm

\noindent {\large {\bf II. The model and simulations:}} 

The Hamiltonian representing the Ising feromagnet 
(described on a square lattice of linear size $L$) in the presence of
a magnetic field is
\begin{equation}
H=-J\sum_{<ij>} {S^z_i}{S^z_j} - \sum_i h(r,t)S^z_i.
\end{equation}
\noindent Where, $S^z_i (=\pm 1)$ is the Ising spins, $J$ is the ferromagnetic
($J > 0$) interaction strength and
$h(r,t)$ is the space and time dependent magnetic field.
The spatially spreading magnetic field is taken as
\begin{equation}
h(r,t)=-h_0 {\rm exp}(-{{((x-x_0)^2+(y-y_0)^2)} \over {2(a^2+b^2t^2)}}).
\end{equation}
\noindent The field $h(r,t)$ is acting along -z direction and spreading
in time ($t$) radially
($r^2=(x-x_0)^2+(y-y_0)^2$) 
 outward from the central site ($x_0, y_0$). 
This is like a unnormalised Gaussian, spreading radially keeping its height
($h_0$) fixed. The factor $h_0$ is defined as the strength of the field. 
Hence, $h_0$ is the magnitude of the spreading field (measured in the unit of $J$)
at the central site for ever.
The value of the field decreases from that at the central site
($x_0, y_0$) as one go radially away from the origin. The effective radius
($r_{\rm eff}$) is defined as that
value of the radius where the value of the field becomes ${1 \over e}$ times
the value at origin. So, $r_{\rm eff} = \sqrt{(2(a^2+b^2t^2)}$.
At time $t=0$ the effective radius was $\sqrt{(2a^2)}$.
As time passes the effective radius $r_{\rm eff}$ increases and eventually
the field spreads over the whole lattice.
The nearest neighbour ferromagnetic uniform ($J=1$) interaction
is considered here. 
The boundary condition is periodic in both sides of the square lattice.

A square of linear size $L = (101)$ is considered and the origin of
spreading field $h(r,t)$ is taken at the center
($x_0=51, y_0=51)$ of the lattice.  
Initially (at $t=0$) value of  all spins ($S^z_i$) are
taken up (+1). In the random spin updating scheme. a site is chosen 
randomly and the energy of the system is calculated (by using equation (1)).
The probability of flip of this chosen spin is obtained from Metropolis
rate\cite{binder}
\begin{equation}
P(S^z_i \to -S^z_i) = {\rm Min}[{\rm exp}(-{{\Delta H} \over {{k_B}T}}),1]
\end{equation}
\noindent where, $k_B$ is Boltzmann constant and $T$ denotes the
temperature (measured in the unit
of $J/{k_B}$).
$\Delta H$ is the change in energy due to spin flip. Using Monte Carlo
method, this spin flipped ($S^z_i \to -S^z_i$)  if the value of
Metropolis rate is less than or equal to a fraction ($0<f<1$ and distributed
uniformly) chosen
randomly. $L^2$ such random updates of spins constitutes one Monte
Carlo Step per Spin (MCSS) and acts as the unit of time in the
simulation. Here, $h(r,t)$ is updated after each MCSS.

\vskip 1cm

\noindent {\large {\bf III. Results:}} 

The Monte Carlo simulation is performed on a two dimensional lattice of linear
size $L$. Here $L=101$, so that the site (51,51) becomes the central site. The
parameters $a$ and $b$ of the
the spreading field were kept
constant ($a=1.0$ and $b=0.1$) throughout the simulational study. Later, the field was allowed
to spread according to the equation (2) and the spins were updated (randomly) with
probability (given in equation -3).
The total magnetisation $m(t)$ is calculated as $m(t) = {{\sum S^z_i} \over {L^2}}$
and if its value becomes less than or equal to 90 percent  of its initial value
, it nucleates and the time required for this is called the nucleation time $\tau_{\rm nucl}$.
This prescription of calculating the nucleation time was used earlier \cite{mads}.

It was observed that a single compact cluster of down spins, embedded in the sea up spins, grows from 
central region.
A typical growth of a cluster of down spins is shown in figure-1. 
Here, $T=0.8$ and $h_0=1.5$ are chosen.
This resuls is quite
obvious here, since the field is spreading radially. As time passes the boundary sites
of negative spin cluster experiences larger 
(than that at earlier time)
values of negative field and get flipped.
In this way the cluster of down spins grows. 
From figure-1(a) and figure-1(b), it is also observed that, in earlier time, the topology
of the lattice is not reflected in the growing clusters of negative spins. 
However, for later time, as the field spreads radially, this topological symmetry of
the lattice starts to appear from the shape of the growing cluster.
It may be noted that
figure-1(c) and figure-1(d) show the clusters of nearly square shaped.

The growth and the nucleation studied here (in the presence of a magnetic field
spatially spreading in time) is compared quantitatively with that in the presence of a 
uniform and static magnetic field. 
The nucleation time was calculated as averaged over 20 different random samples.
In this case one expects that for higher 
values of field strength the nucleation time (for spreading field) would be
many times higher than that for uniform and static field. 
The reason is very clear here. In this case all spins will not experience 
the maximum amount of field at any given instant (MCSS).
However, this time (for spreading field)
would become of the same order of that for uniform and static field as one apply the
low field. 
Here, the nucleation time is quite large for the case of static field. And 
for spreading field, the field will get sufficient time to spread over the whole 
lattice. As a result, the nucleation time ($\tau_{nucl}$) for spreading field will be of the same
order of that of static field.
The total number of downspins, $N^{\downarrow}_{\rm eff}$,  within the circle of effective radius 
($r_{\rm eff}$) is also calculated. Hence, the magnitude of effective magnetisation,
($m_{\rm eff} (= {{N^{\downarrow}_{\rm eff}} \over {N_{\rm total}}})$, is calculated. 
$N_{\rm total}$ is equal to the total number of spins, i.e., $L^2$.
As time passes,
the $m_{\rm eff}$ increases. 
A new time scale, called effective time ($\tau_{eff}$), may be defined here as follows:
the time required by the magnitude of the effective magnetisation to reach the value
0.01. This effective time is a time associated to the developement of local magnetisation
within the circle of radius $r_{eff}$. 

These results are depicted in figure-2. This study clearly indicates the difference
in time of the growth and nucleation for different types of fields (uniform-static type 
and spreading type).

The growth of effective magnetisation ($m_{\rm eff}$) at a fixed
temperature ($T=0.80$) and for the different values of the strength of the magnetic field ($h_0=0.60, 0.65,
 0.80$) are
studied. These are shown in figure-3. For any fixed value of field strength, the growth $m_{\rm eff}$
is very slow and suddenly it starts to grow very rapidly. 
Naturally,
the $\tau_{\rm eff}$ will increase for smaller values of the strength of spreading magnetic field
at any given temperature of the system.
This feature is clear from figure-3. 

The growth of $m_{\rm eff}$ is also studied for a fixed value of field strength ($h_0=0.8)$ and different
values of temperature ($T$=0.8, 1.0 and 1.2). This is shown in figure-4. Here, also it is observed that
the effective time ($\tau_{\rm eff}$) increases as the temperature decreases, for a fixed value of the
maximum magnitude of the spreading magnetic field. It may be mentioned here, that in the
case of nucleation by static field, at $T=0.8$, $h_0=1.5$ would be in the multidroplet
region and $h_0=0.8$ would be near the crossover of multidroplet and single-droplet region for
the system size considered here \cite{rik0}.

To consider the effective time ($\tau_{\rm eff}$) as a new time scale, in the case of nucleation
in Ising ferromagnet, would be justified in the following study. This is the main part of this paper.
The ratio ($R={{\tau_{\rm nucl}} \over {\tau_{\rm eff}}}$) of nucleation time 
($\tau_{\rm nucl}$) to the effective time ($\tau_{\rm eff}$) is studied
as a function of the strength of spreading magnetic field for different values of temperature.
This is shown in figure-5. The ratio ($R$) shows a {\it nonmonotonic} variation with the strength of
spreading magnetic field. The reason behind this nonmonotonic variation of $R$ with $h_0$, is due
to the growth of the clusters formed by the negetive spins, which happens only within the circle of effective radius
,i.e., $r_{\rm eff}$. The lifetime of metastable state, i.e., $\tau_{\rm nucl}$, is solely dependent on the time
of growth of negative clusters of a required value of magnetisation. However, in the present study, this is solely
governed by a single compact cluster of negative spins which resides within the circle of effective
radius ($r_{\rm eff}$). The rate of spreading (${{dh(t)} \over {dt}}$) of the field, is a function of time and the radius ($r_{\rm eff}$)
of the effective circle. So, it is expected that the nucleation time ($\tau_{\rm nucl}$) and the effective time ($\tau_{\rm eff}$) have
different types of dependence on $h_0$. Hence, $R$ would show a nonmonotonic (nonlinear) dependence on $h_0$.  

As the field increases, the ratio increases first, and then reaches a maximum,
then it starts to decrease.
Since the rate of spreading of the field, is not constant, there may exist some value of the field, for
which the nucleation time ($\tau_{\rm nucl}$), differs widely, from effective time ($\tau_{\rm eff}$).
Here, in this region near the peak, the growth of $m_{\rm eff}$ becomes much slower than the growh of $r_{\rm eff}$.
Hence, $\tau_{\rm nucl}$ becomes much larger than $\tau_{\rm eff}$ and consequently $R$ becomes maximum.
This is a possible qualitative explanation of getting a peak of $R$ for a definite value of $h_0$. However,
further extensive simulational investigation is required to have some quantitative idea about the
specific value of $h_0$ which maximises $R$. Precisely, the $R$ takes the value, depending on
$h_0$, appears to be the competition of rates of growth of $m_{eff}$ and the magnetisation for
reversal. The maximum of $R$, is that value, where the nuclation time is relatively high for a
particular value of $h_0$. This has an experimental significance also. The reversal or switching
time may be obtained accordingly if the amplitude of the spreading field is suitably adjusted.
This may be used in magnetic storage device to have an optimum condition of speed of recording
and the longivity of the recorded media.

 This indicates that the effective time ($\tau_{\rm eff}$) may be considered
as to define a {\it new scale of time} in the problem of nucleation in Ising ferromagnet in the presence of a
{\it spreading} magnetic field. 
If the effective time would not provide another scale of time, this would be a simple factor of nucleation time. 
Consequently, if the ratio of nucleation time and the effective time was plotted against the applied
field strength, it would show a straight line parallel to the horizontal axis. 
However, in this case, $R$ shows a nonmonotonic variation with $h_0$.
The maximum value of the ratio 
($R_{\rm max}$ and the strength of the field
($h_0^{\rm max}$) for which $R$ becomes maximum, 
is plotted against the temperature ($T$) and shown in
figure-6. Both $h_0^{\rm max}$ and $R_{\rm max}$ decreases nonlinearly as the temperature
increases.

\vskip 1cm

\noindent {\large {\bf IV. Summary:}} 

In this paper, the nucleation in Ising ferromagnet is studied by Monte Carlo simulation using Metropolis
single spin flip dynamics.  All earlier studies are performed with magnetic field uniformly distributed over the
space and constant in time. A few studies are done by using a slowly varying (in time) magnetic
field but uniform over the space. {\it However, the present study differs significantly from all other
earlier studies}. Here, the applied magnetic field has a {\it spatio-temporal} variation. The applied
magnetic field is spreading over the space in time. A Gaussian-like simple form of spreading magnetic field is
taken here. The width of the Gaussian is spreading in time keeping the altitute fixed.  

The nucleation time would become larger (for fixed values of temperature and field strength) in
comparison to that observed in the case of static and uniform magnetic field. In the case of
static and uniform magnetic field, the nucleation time provides the scale of time. A considerable
number of studies are done using this time scale which was also considered as the lifetime of the
metastable state. However, in the case of spreading magnetic field,
apart from the nucleation time, a new scale of time appears. This is the effective time. 
Interestingly, it 
is observed in the
present study, the variation of this ratio with field strength is {\it nonmonotonic}. Hence, one would say
that the effective time and nucleation time provide separately different scales of time in this
problem of nucleation in spreading magnetic field.

The effective time and the nucleation time are studied here as a function of the factor 
predominantly determining the rate
of spreading of the field. It is observed that both 
show similar power law type decay.         

The study of the nucleation in the presence of a spreading magnetic field would help to 
sustain the metastability for a longer period if one uses the magnetic field is spreading (in time) very slowly. 
In the field of technology of magnetic recording,
this may help to increase the longivity of recorded media, if the rate of spreading is 
suitably managed. 
Hopefully, this study will also explore some
new class of problems in nonequilibrium statistical mechanics in near future.

\vskip 1cm

\noindent {\bf Acknowledgments:} The library facility provided by the
Central Library of Calcutta University is gratefully acknowledged. Author
would like to thank D. Stauffer for important discussions.

\newpage
\setlength{\unitlength}{0.240900pt}
\ifx\plotpoint\undefined\newsavebox{\plotpoint}\fi
\sbox{\plotpoint}{\rule[-0.200pt]{0.400pt}{0.400pt}}%


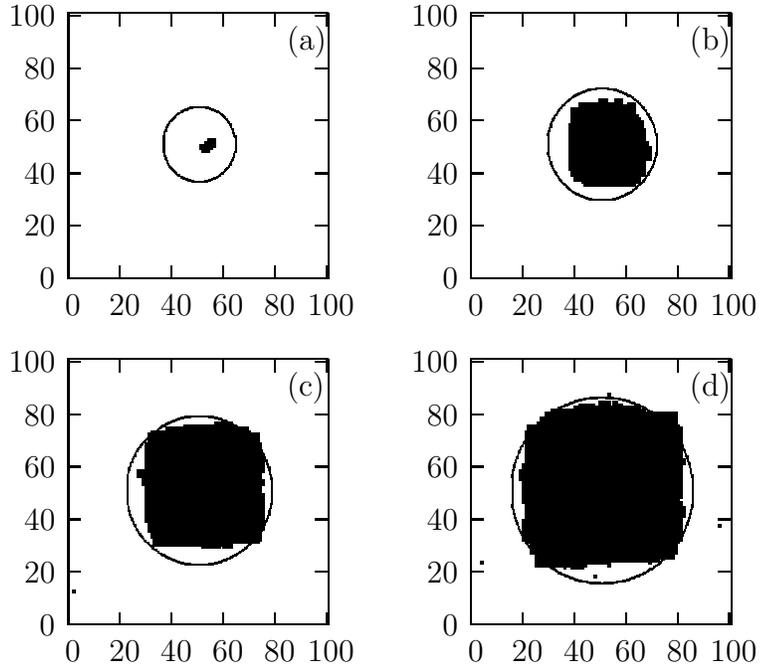
\noindent {\bf Fig.1} Growth of negative spin clusters (marked by black dots)
(on a $101\times101$ square lattice) with time ($t$). The solid line is a circle
of radius $r_{eff}$ in each diagram.
Here, $T=0.8$ and $h_0=1.5$. (a) $t=100$ MCSS, (b) $t=150$ MCSS,
(c) $t=200$ MCSS and (d) $t=250$ MCSS. Here, $a=1.0$ and $b=0.1$.
\newpage

\setlength{\unitlength}{0.240900pt}
\ifx\plotpoint\undefined\newsavebox{\plotpoint}\fi
\sbox{\plotpoint}{\rule[-0.200pt]{0.400pt}{0.400pt}}%
%

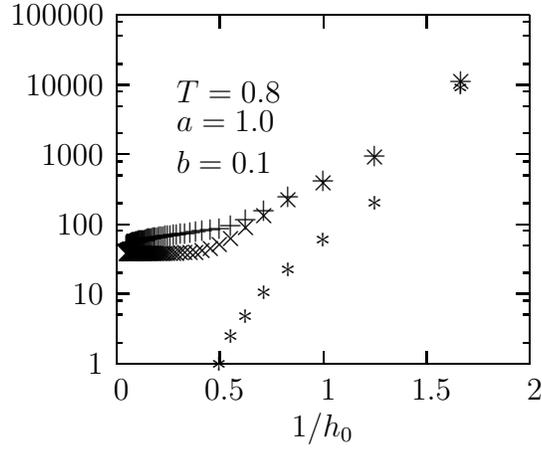
\noindent {\bf Fig-2.} Dependences of $\tau_{\rm nucl}(+)$, $\tau_{\rm eff}(\times)$ with the amplitude ($h_0$) of spreading
field and that of $\tau_{\rm nucl}(\ast)$ with the magnitude ($h_0$) of static uniform field. Here, $T=0.8$, $a=1.0$ and
$b=0.1$.

\newpage

\setlength{\unitlength}{0.240900pt}
\ifx\plotpoint\undefined\newsavebox{\plotpoint}\fi
\sbox{\plotpoint}{\rule[-0.200pt]{0.400pt}{0.400pt}}%
%

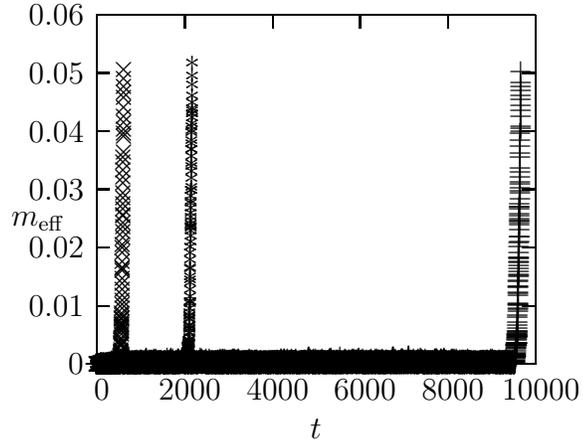
\noindent {\bf Fig-3.} The effective magnetisation ($m_{\rm eff}$) is plotted against the time ($t$). Different
symbols corresponds to different values of amplitudes ($h_0$) of the spreading field. $h_0=0.60(+)$,
$h_0=0.65(\ast)$ and $h_0=0.80(\times)$. Here, in all cases $T=0.8$, $a=1.0$ and $b=0.1$.

\newpage

\setlength{\unitlength}{0.240900pt}
\ifx\plotpoint\undefined\newsavebox{\plotpoint}\fi
\sbox{\plotpoint}{\rule[-0.200pt]{0.400pt}{0.400pt}}%
%

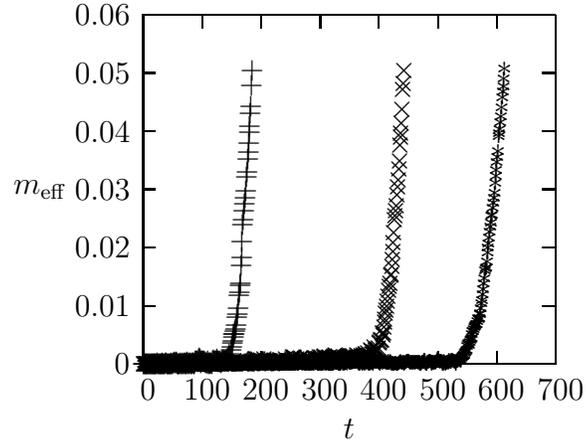
\noindent {\bf Fig-4.} The effective magnetisation ($m_{\rm eff}$) plotted
against time ($t$). Different symbols corresponds to different values of
temperatures ($T$). $T = 1.2(+)$, $T = 1.0 (\times)$ and
$T = 0.8 (\ast)$. Here, the value of field strength $h_0 = 0.8$, $a=1.0$ and $b=0.1$.

\newpage

\setlength{\unitlength}{0.240900pt}
\ifx\plotpoint\undefined\newsavebox{\plotpoint}\fi
\sbox{\plotpoint}{\rule[-0.200pt]{0.400pt}{0.400pt}}%
%

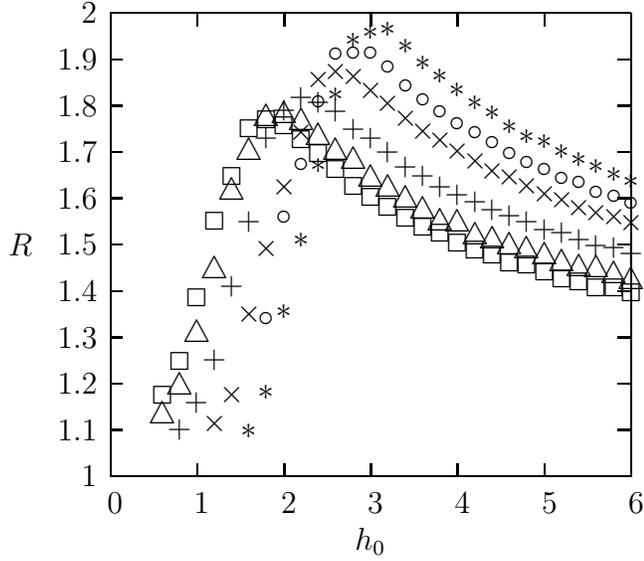
\noindent {\bf Fig-5.} The ratio ($R = {{\tau_{\rm nucl}} \over {\tau_{\rm eff}}}$) is plotted against the field
strength ($h_0$) for different temperatures ($T$). Different symbols
corresponds to different temperatures. $T=1.3 (\Box)$, $T=1.2 (\triangle)$, 
$T=1.0 (+)$, $T=0.8 (\times)$, $T=0.7 (\circ)$ and $T=0.6 (\ast)$. Here, $a=1.0$ and $b=0.1$.
\newpage

\setlength{\unitlength}{0.240900pt}
\ifx\plotpoint\undefined\newsavebox{\plotpoint}\fi
\sbox{\plotpoint}{\rule[-0.200pt]{0.400pt}{0.400pt}}%
%

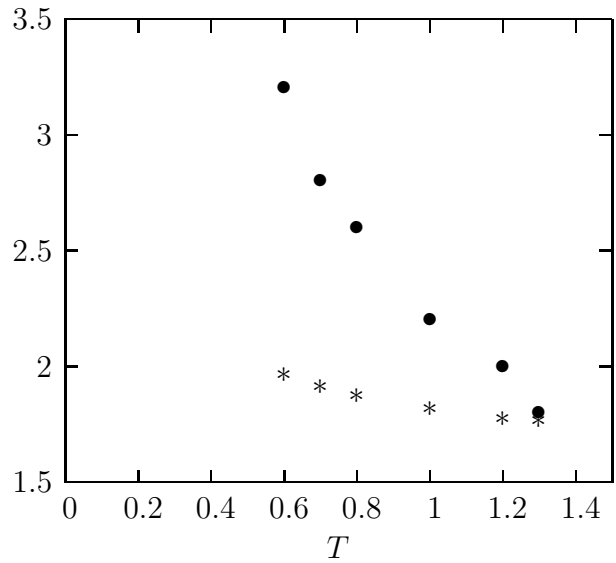
\noindent {\bf Fig-6.} The $h_0^{\rm max} (\bullet)$ and 
$R_{\rm max} (\ast)$ plotted against the temperature ($T$).
Here, $a=1.0$ and $b=0.1$.

\end{document}